\documentclass[fignum]{apa}
\usepackage{amsfonts}
\usepackage{amsmath}
\usepackage{graphicx}
\usepackage[active]{srcltx} 
\usepackage{natbib}
\usepackage{graphicx,bm,amsmath}

\title{Estimation of latent variable models for ordinal data via fully exponential Laplace approximation}
\author{Silvia Bianconcini and Silvia Cagnone}
\affiliation{Department of Statistics, University of Bologna.}
\shorttitle{Approximate inference in latent variable models}

\begin{document}
\maketitle

\abstract{Latent variable models for ordinal data represent a useful
tool in different fields of research in which the constructs of
interest are not directly observable. In such models, problems
related to the integration of the likelihood function can arise
since analytical solutions do not exist. Numerical approximations,
like the widely used Gauss Hermite (GH) quadrature, are generally
applied to solve these problems. However, GH becomes unfeasible as
the number of latent variables increases. Thus, alternative
solutions have to be found. In this paper, we propose an extended
version of the Laplace method for approximating the integrals, known
as fully exponential Laplace approximation. It is computational
feasible also in presence of many latent variables, and it is more
accurate than the classical Laplace method.}


\section{Introduction}
Latent variable models represent a useful tool in the social sciences where the analyzed constructs cannot be directly observed and, hence, they are not measurable. However, a set of indicators related to each unobserved variable can be measured. They are often coded into a number of ordered categories, so that latent variable models with ordinal variables have to be used. These models can be defined within the Generalized Linear Latent Variable Model (GLLVM)  framework \citep{BartKnot99,MouKno01}, according to which the entire set of the responses given by an individual to a certain number of items, called response pattern, is expressed as a function of one or more latent variables through a monotone differentiable link function.  The estimation of the model parameters can be obtained by means of a full information maximum likelihood method via the EM algorithm, that guarantees quite accurate estimates \citep{Mou00, Mou03}.\\ The presence of the latent variables causes problems related to the integration of the likelihood function, since analytical solutions do not exist. In order to overcome this drawback, numerical approximations are usually applied. One of the most often used technique is the classical Gauss-Hermite (GH)  quadrature \citep{BoAit81}, that provides quite good parameter estimates when many quadrature points are considered per each latent variable. However, it becomes computational unfeasible as the number of latent variables increases. This represents a serious limitation for a large number of applications where several observed and latent variables are required \citep{CaMouVa09}.\\
As alternative solution to GH, the Adaptive Gauss Hermite (AGH)
quadrature has been discussed for different models with random
effects and/or latent variables \citep{PiBa95,HesSkPi05,SchBo05}. In
all these studies, AGH is shown to perform better than GH, also when
few quadrature points are used. Indeed, it consists of adjusting the
GH nodes with the first and second moments of the posterior density
of the latent factors given the manifest variables. This allows a better approximation of the function to be integrated. Nevertheless, the AGH is very computational intensive, particularly in latent variable models for ordinal data \citep{CaMo11}. \\
An approximation technique that is not affected by the presence of high dimensional integrals is the Laplace method
\citep{DeBr81,BarCox89}, that can be viewed as a particular case of the AGH when just one abscissa is used \citep{LiuPie94}.
Given its reduced dimensionality, the Laplace method is one of the fastest technique, since the computational burden depends only on the calculation of the mode of the integrand \citep{TieKa86,Rade00,HubRonFe04,PinChao06}.  However, the Laplace approximation has an error of order $O(p^{-1})$, that depends only on the number of items $p$, hence it is not directly controllable.
Moreover, \cite{Joe08}  has investigated its performance for a variety of discrete response mixed  models, and he has found that it becomes less adequate as the degree of discreteness increases.\\
When either the EM algorithm or a direct maximization of the observed data log-likelihood is used for model estimation, an extended version of the Laplace method, called Fully exponential Laplace Approximation (FLA), can be applied. It has been introduced and developed by \cite{TieKa89}  in the Bayesian context for approximating posterior distributions. Recently, it has been extended by \cite{RizVer09} to a variety of models for longitudinal continuous measurements and time-to-event data estimated via the EM algorithm. The main idea proposed by these authors is to apply the FLA to the expected score function of the model
parameters with respect to the posterior distribution of the latent variables. With the FLA, a better approximation of the multidimensional integrals is achieved,
being the approximation error of order $O(p^{-2})$. Moreover, the computational complexity of this approach is similar to the classical Laplace method since it depends only on the numerical optimization required to compute the mode of the integrand.\\
In this paper, we extend the FLA for the general class of latent
variable models for ordinal data within the GLLVM context. In
Section 2, the models for ordinal data are introduced, whereas in
Section 3 the estimation problem is discussed, with particular
attention to the fully exponential Laplace approximation. In Section
4, a simulation study is performed in order to compare the finite
sample and asymptotic properties of the AGH and FLA under different
conditions. Finally, Section 5 gives the conclusions.

\section{Model specification}
Let $\mathbf{y}$ be a vector of $p$ ordinal observed variables each of them with $c_{i}$ categories, and $\mathbf{z}$ be a vector of $q$
latent variables. The $c_{i}$ \mbox{($i=1,\ldots,p$}) ordered categories of the variables $y_{i}$ have associated the probabilities
$\pi_{i,1}(\mathbf{z}),\pi_{i,2}(\boldsymbol z),\ldots,\pi_{i,c_{i}}(\mathbf{z})$, which are functions of the vector of the latent variables $\mathbf{z}$.\\ Following the general scheme of the GLLVM framework, the probability associated to $\mathbf{y}$ is given by
\begin{equation} \label{eq:1.1}
f(\mathbf{y})=\int_{R^q}
g(\mathbf{y}|\mathbf{z})h(\mathbf{z})\textit{d}\mathbf{z}
\end{equation}
where $h(\mathbf{z})$ is assumed to be a multivariate standard normal
distribution. $g(\mathbf{y}|\mathbf{z})$ is the conditional probability of
the observed variables given $\mathbf{z}$. It is assumed to follow a
multinomial distribution
\begin{equation} \label{cond}
g(\mathbf{y}\mid \mathbf{z})=\prod_{i=1}^{p}g(y_{i}\mid
\mathbf{z})
\end{equation}
where
\begin{equation} \label{cond1}
g(y_{i}\mid \mathbf{z})=\prod_{s=1}^{c_{i}-1}\left(\frac{\gamma_{i,s}(\mathbf{z})}{\gamma_{i,s+1}(\mathbf{z})-\gamma_{i,s}(\mathbf{z})}\right)^{y_{i,s}^{*}}\left(\frac{\gamma_{i,s+1}(\mathbf{z})-\gamma_{i,s}(\mathbf{z})}{\gamma_{i,s+1}(\mathbf{z})}\right)^{y_{i,s+1}^{*}-y_{i,s}^{*}}.
\end{equation}
Expression (\ref{cond}) is obtained by assuming the conditional independence of the observed variables given the latent variables. In expression (\ref{cond1}),
$\gamma_{i,s}(\mathbf{z})=\pi_{i,1}(\mathbf{z})+\pi_{i,2}(\mathbf{z})+\ldots+\pi_{i,s}(\mathbf{z})$ is the probability of a response in category $s$ or lower on the variable $i$, and it is function of $\mathbf{z}$. For simplicity, from now on we consider $\gamma_{i,s}=\gamma_{i,s}(\mathbf{z})$. $y_{i,s}^{*}$ is equal to 1 if the response $y_{i}$ is in the category $s$ or lower, and 0 otherwise. \\
As in the classical generalized linear model, the systematic component  is
defined as
\begin{equation} \label{eta}
\eta_{i,s}=\tau_{i,s}-\sum_{j=1}^q\alpha_{ij}z_{j},\quad
s=1,\ldots,c_{i}-1, i=1,\ldots,p
\end{equation}
where $\eta_{i,s}$ is the linear predictor, and $\tau_{i,s}$ and $\alpha_{ij}$ can be interpreted as thresholds and factor loadings of the model. For the thresholds, the inequality  $-\infty=\tau_{i,0}\leq \tau_{i,1}\leq \tau_{i,2}\leq\ldots\leq \tau_{i,c_{i}}=+\infty$ holds. Each factor loading $\alpha_{ij}$ measures the effect of the correspondent latent variable $z_{j}$ on some function of the cumulative probability $\gamma_{i,s}$. \\
The relation between the systematic component and the conditional
means of the random component distributions is given by
$\eta_{i,s}=\nu_{i,s}(\gamma_{i,s})$, where $\nu_{i,s}$ is the link
function and can be any monotonic differentiable function. Here, we
refer to the logit link function, so that eq. $(\ref{eta})$ is known
as proportional odds model. However, other link functions can be
chosen.

\section{Model estimation}
Model estimation is achieved by using the maximum likelihood through the EM algorithm, since the latent variables are unknown. At this regard, we apply a full information maximum likelihood method by which all the parameters of the model are estimated simultaneously.\\ For a random sample of size $n$, from equation  (\ref{eq:1.1}), the observed data log-likelihood is defined as
\begin{equation} \label{loglike}
L=\sum_{l=1}^{n} \log f({\bf y}_l) =
\sum_{l=1}^{n} \log \int_{R^{q}}g({\bf{y}}_{l} \mid {\bf z}_{l})
h({\bf z}_{l})d {\bf z}_{l}.
\end{equation}
The EM algorithm consists of an Expectation step (E-step), in which the expected score function  $E(S(\mathbf{a}_{i}))$ of the model parameters $\mathbf{a}'_{i}=(\tau_{i,1},\ldots\tau_{i,c_{i}-1}, \alpha_{i1},\ldots, \alpha_{iq}), i=1,\cdots,p$,
is computed. The expectation is with respect to the posterior distribution $h({\bf z}_{l} \mid {\bf y}_l)$ of ${\bf z}$ given the observations for each individual. In the Maximization step (M-step), updated parameter estimates are obtained by equating to 0 the expected score functions.\\
\cite{Lui82} proved that maximizing the observed data score vector $\partial L /\partial \mathbf{a}_{i}$ is equivalent to maximize the expected score function $E(S(\mathbf{a}_{i}))$ with respect to $h({\bf z}_{l} \mid {\bf y}_l)$,  so that
\begin{eqnarray} \label{score}
\frac{\partial L}{\partial \mathbf{a}_{i}}=E(S(\mathbf{a}_{i}))&=&\sum_{l=1}^{n} \frac{\int S(\mathbf{a}_{i})g(\mathbf{y}_{l}\mid\mathbf{z}_{l})h(\mathbf{z}_{l})d\mathbf{z}_{l}}{\int g(\mathbf{y}_{l}\mid\mathbf{z}_{l})h(\mathbf{z}_{l})d\mathbf{z}_{l}}= \\ \nonumber
&=&\sum_{l=1}^{n} \int \frac{\partial \log g({\bf{y}}_{l} \mid {\bf z}_{l})}{\partial \mathbf{a}_{i}} h({\bf z}_{l} \mid {\bf y}_l)d{\bf z}_{l}  \quad i=1,\ldots,p
\end{eqnarray}
where
\begin{eqnarray} \label{der}
S(\mathbf{a}_{i})=\frac{\partial \log g({\bf{y}}_{l} \mid {\bf z}_{l})}{\partial \mathbf{a}_{i}}&=&\sum_{s=1}^{c_{i}-1}\left[ y_{i,s,l}^{*}\frac{\partial\theta_{i,s,l}(\mathbf{z}_{l})}{\partial\mathbf{a}_{i}} -y_{i,s+1,l}^{*}\frac{\partial b(\theta_{i,s,l}(\mathbf{z}_{l}))}{\partial\mathbf{a}_{i}}\right] \nonumber
\end{eqnarray}
and
\begin{equation} \label{theta}
\theta_{i,s,l}(\mathbf{z}_{l})=\log \frac{\gamma_{i,s,l}}{(\gamma_{i,s+1,l}-\gamma_{i,s,l})}  \quad s=1,\ldots,c_{i}-1, i=1,\ldots, p.
\end{equation}
\begin{equation}  \label{beta}
b(\theta_{i,s,l}(\mathbf{z}_{l}))=\log \frac{\gamma_{i,s+1,l}}{(\gamma_{i,s+1,l}-\gamma_{i,s,l})} \quad s=1,\ldots,c_{i}-1, i=1,\ldots, p.
\end{equation}
The expressions of the derivatives reported in the last equality of (\ref{der})  with respect to thresholds and loadings  can be found in \cite{Mou00, Mou03}.\\
From  eq. (\ref{score}), it can be noticed  that the computation of the expected score functions involves a multidimensional integral that  cannot be solved analytically, hence numerical approximations are required. In particular, in  the following,  we propose the use of an extended version of the classical Laplace approximation, that is the fully exponential Laplace method.

\subsection{Fully exponential Laplace approximation method}
The FLA method has been proposed for the first time by \cite{TieKa89} in order to approximate posterior distributions in the Bayesian context. It represents an extension of the classical Laplace approximation that, as known, is based on the second order Taylor expansion of the logarithm of the integrand, with the latent variables evaluated at the mode (see, among the others, \cite{TieKa86}).\\ The Laplace method has the advantage of dealing with integrals of any dimensionality without introducing computational problems but, for the general class of latent variable models discussed in this paper, it produces an approximation error of  order $O(p^{-1})$, that can be reduced only increasing the number of observed variables.
The FLA leads to an improvement of the approximation error maintaining the same computational complexity as the classical Laplace method. The extension of FLA to joint models for continuous longitudinal measurements and time-to-event data has been proposed by \cite{RizVer09}. It requires the computation of the following quantities
\begin{equation} \label{Acomp}
E(A(\mathbf{z}_{l}))=\int A(\mathbf{z}_{l})h(\mathbf{z}_{l}\mid\mathbf{y}_{l})d\mathbf{z}_{l}
=\frac{\int A(\mathbf{z}_{l})g(\mathbf{y}_{l}\mid\mathbf{z}_{l})h(\mathbf{z}_{l})d\mathbf{z}_{l}}{\int g(\mathbf{y}_{l}\mid\mathbf{z}_{l})h(\mathbf{z}_{l})d\mathbf{z}_{l}}
\end{equation}
that differ from (\ref{score}) since $A(\cdot)$ are the components of the score functions $S(\mathbf{a}_{i})$ that depend on the latent variables.\\
The main idea of FLA is to approximate both the numerator and the denominator in eq. (\ref{Acomp}) with the classical Laplace method. \cite{TieKa86} proved that the error terms of  order $O(p^{-1})$ in the numerator and the denominator cancel out, leading to a smaller error term of order $O(p^{-2})$.\par
To extend the FLA to the proportional odds model discussed in this paper, we have
to take into account for the derivatives (\ref{der}) with respect to the thresholds and the loadings,
that are characterized by different  $A(\mathbf{z}_{l})$ components. In more detail, from the derivatives of the logarithm
of $ g(y_{i,l} \mid {\bf z}_{l})$ with respect to the thresholds we get
$$A_{1,i,s}(\mathbf{z}_{l})=-\frac{\partial \theta_{i,s-1,l}(\mathbf{z}_{l})}{\partial\tau_{i,s}}=\left\{
                                                                                            \begin{array}{ll}
                                                                                              (1-\gamma_{i,s,l}), & \hbox{if $s=1$;} \\
                                                                                              (1-\gamma_{i,s,l})\frac{\gamma_{i,s,l}}{(\gamma_{i,s,l}-\gamma_{i,s-1,l})}, & \hbox{if $s=2,\ldots,c_{i}-1$.}
                                                                                            \end{array}
                                                                                          \right.
$$\\
$$A_{2,i,s}(\mathbf{z}_{l})=\frac{\partial b(\theta_{i,s,l}(\mathbf{z}_{l}))}{\partial\tau_{i,s}}
=(1-\gamma_{i,s,l})\frac{\gamma_{i,s,l}}{(\gamma_{i,s+1,l}-\gamma_{i,s,l})} \quad s=1,\ldots,c_{i}-1.$$\\
From the derivatives of the logarithm of $g(y_{i,l} \mid {\bf z}_{l})$ with respect to the loadings $\boldsymbol \alpha_{i}=(\alpha_{i1},...,\alpha_{iq})$, we get
$$A_{3,i,s}(\mathbf{z}_{l})=-\frac{\partial\log g(y_{i,s,l}\mid\mathbf{z}_{l})}{\partial\boldsymbol\alpha_{i}}
=\left\{
                                                                                            \begin{array}{ll}
                                                                                             (1-\gamma_{i,s,l}) \mathbf{z}_{l}, & \hbox{if $s=1$;} \\
                                                                                             (1-\gamma_{i,s,l}-\gamma_{i,s-1,l})\mathbf{z}_{l}, & \hbox{if $s=2,\ldots,c_{i}$.}
                                                                                            \end{array}
                                                                                          \right.
$$

The FLA approximation can be applied only to strictly positive
functions $A(\cdot)$. In our case, this condition is not necessarily
guaranteed since the  $A(\cdot)$ are components of the score
functions, not constrained to be positive. To overcome this
problem,  the method of the moment generating function can be used.
According to this approach, since the quantity
$\exp\{t'A(\mathbf{z}_{l})\}$ is always positive, the FLA
approximation can be applied to the moment generating function
$M(t)=E[\exp\{t'A(\mathbf{z}_{l})) \}$, with latent variables
$\mathbf{z}_{l}$ evaluated at the mode $\hat{\mathbf{z}}_{l}=\arg
\max _{\mathbf{z}_{l}}[\log g(\mathbf{y}_{l}\mid\mathbf{z}_{l})
+\log h(\mathbf{z}_{l})+ t'A(\mathbf{z}_{l})]$. In doing so, we get
the approximate moment generating function $\hat{M}(t)$.  Hence,
from the corresponding cumulant-generating function $\log
\hat{M}(t)$, we obtain the approximate expected values
$\hat{E}(A(\mathbf{z}_{l}))$. These latter are the quantities of
interest, and they are given by
\begin{equation} \label{cumgf}
\hat{E}(A(\mathbf{z}_{l}))=\frac{\partial}{\partial t} \log \hat{M}(t) |_{t=0} = \frac{\partial}{\partial t}  \log \hat{E}[\exp\{t'A(\mathbf{z}_{l})) \}]\Big|_{t=0}.
\end{equation}
\cite{TieKa89} proved (Theorem 2, pag. 712) that eq. (\ref{cumgf}) is equivalent to the following expression
\begin{eqnarray}\label{ExpA}
\hat{E}(A(\mathbf{z}_{l}))&=& A(\mathbf{\hat{z}}_{l}) + \frac{\partial \log\det (\boldsymbol\Sigma_{l}^{(t)})^{-1/2}}{\partial t}\Big|_{(\mathbf{z}_{l}=\mathbf{\hat{z}}_{l}, t=0)}+ O(p^{-2})= \\ \nonumber
&=& A(\mathbf{\hat{z}}_{l}) - \frac{1}{2} \mathrm{tr}(\boldsymbol\Omega)\Big|_{(\mathbf{z}_{l}=\mathbf{\hat{z}}_{l}, t=0)} + O(p^{-2})
\end{eqnarray}
where
\begin{eqnarray} \label{Sigma}
\boldsymbol\Sigma_{l}^{(t)} &=& -\frac{\partial^{2} \{\log g(\mathbf{y}_{l}\mid\mathbf{z}_{l})+\log h(\mathbf{z}_{l})+t' A(\mathbf{z}_{l}) \}}{\partial \mathbf{z}_{l}' \partial \mathbf{z}_{l}}= \\ \nonumber
&=& \sum_{i=1}^{p}\sum_{s=1}^{c_{i}-1}\left\{\boldsymbol \alpha_{i}\boldsymbol \alpha_{i}'\left[-y_{i,s,l}^{*}\gamma_{i,s+1,l}(1-\gamma_{i,s+1,l})+y_{i,s+1,l}^{*}\gamma_{i,s,l}(1-\gamma_{i,s,l})\right]\right\}+\boldsymbol I-\frac{\partial^{2} t'A(\mathbf{z}_{l})}{\partial \mathbf{z}_{l}' \partial \mathbf{z}_{l}}
\end{eqnarray}
and
$$\boldsymbol\Omega=\boldsymbol\Sigma_{l}^{-1}\{\partial\boldsymbol\Sigma_{l}^{(t)}/\partial t \}.$$
The expressions of the first derivatives of $\boldsymbol\Sigma_{l}$ with respect to $t$ are reported in the Appendix.
\subsection{EM algorithm}
\noindent The steps of the EM algorithm are defined as follows:\\
\begin{enumerate}
\item \noindent Choose initial values for the parameters  $\mathbf{a}'_{i}=(\tau_{i,1},\ldots\tau_{i,c_{i}-1}, \alpha_{i1},\ldots, \alpha_{iq}) \quad i=1,\cdots,p$.

\item \noindent  Compute the mode $\hat{\mathbf{z}}_{l}, l=1,\ldots,n$, by using a Newton Raphson iteration scheme. In more detail, for the $(m)$-th iteration $$\hat{\mathbf{z}}_{l}^{m}=\hat{\mathbf{z}}_{l}^{(m-1)}-[(\boldsymbol\Sigma_{l} ^{(m-1)})^{-1}S({\mathbf{z}}_{l}^{(m-1)})]|_{(\mathbf{z}_{l}=\mathbf{\hat{z}}_{l}^{(m-1)}, t=0)}$$ where $\boldsymbol\Sigma_{l}$ is the Hessian matrix defined in expression (\ref{Sigma}) and $S({\mathbf{z}}_{l})$ is defined as follows
                                 \begin{eqnarray} \label{Scorez}
                                 S({\mathbf{z}}_{l})&=&-\frac{\partial \{\log g(\mathbf{y}_{l}\mid\mathbf{z}_{l})+\log h(\mathbf{z}_{l})+t' A(\mathbf{z}_{l}) \}}{\partial \mathbf{z}_{l}}=\\
                                 &=& \sum_{i=1}^{p}\sum_{s=1}^{c_{i}-1}\left\{\boldsymbol \alpha_{i}'\left[y_{i,s,l}^{*}\gamma_{i,s+1,l}-y_{i,s+1,l}^{*}\gamma_{i,s,l}\right]\right\}+\mathbf{z}_{l}-\frac{\partial t'A(\mathbf{z}_{l})}{\partial \mathbf{z}_{l}}.
                                 \end{eqnarray}

\item \noindent E-step. Compute the FLA expected values $\hat{E}(A_{1,i,s}(\mathbf{z}_{l}))$, $\hat{E}(A_{3,i,s}(\mathbf{z}_{l}))$, for $s=1,\ldots,c_{i}-1, i=1,\ldots,p$,  and $\hat{E}(A_{2,i,s-1}(\mathbf{z}_{l}))$ for  $s=2,\ldots,c_{i},i=1,\ldots,p$, and the approximate expected score function $\hat{E}(S(\mathbf{a}_{i}))$, where $i=1,\ldots,p$.

\item \noindent M-step. Obtain improved estimates for the model parameters $\mathbf{a}'_{i}=(\tau_{i,1},\ldots\tau_{i,c_{i}-1}, \alpha_{i1},\ldots, \alpha_{iq}), \quad i=1,\cdots,p$. For all of them, a Newton Raphson iterative scheme is used in order to solve the corresponding nonlinear maximum likelihood equations.

\item \noindent Repeat steps 2-3-4 until convergence is attained.
\end{enumerate}

\section{Simulation study}
The properties of the FLA method for the proportional odds model can
be evaluated by performing a simulation study in which several
conditions are taken into account. The results will be compared with
those obtained using the AGH quadrature. In recent years, the latter
has been widely applied in latent variable models, since it allows
to obtain estimates that are as accurate as those derived by the GH
technique, but using a small number of quadrature points. It
essentially consists of scaling and translating the classical
Gaussian quadrature locations to place them under the peak of the
integrand, and two different procedures have been adopted in the
literature. According to the first one, the mode of the integrand and the inverse of the information matrix of the integrand evaluated at the mode are computed \citep{LiuPie94,PiBa95,SchBo05}. The advantage of this approach lies in the fact that the quadrature points are not involved in these computations. However, this method is computationally demanding since it requires numerical optimization routines and the computation of second derivatives. Moreover, when parameter estimates are obtained by using iterative algorithms, like in our case, the first and second order moments have to be computed at each step, hence the algorithm becomes very slow.\\An alternative procedure consists of computing the posterior means and  covariance matrices at each step of the algorithm \citep{HesSkPi05}.  Although this method requires the use of quadrature points themselves, the posterior moments should better describe the integrand in those cases in which its tails are heavier than the normal density. In the following, we show how both these techniques work in latent variable models for ordinal data, and we compare their performances with FLA.\\
The softwares used for the analyses are Fortran 95 and R. The codes are available from the authors upon request.

\subsection{Finite sample properties of the estimators}
\begin{center}
\begin{figure}[!htbp]
 \caption{Individual posterior densities with different shapes (on the left side $\beta_{1,2}$=0.000, and on the right side $\beta_{1,2}$=0.168) in the generated population of 200 subjects.\label{Fig1}}
 \hspace{2cm}\rotatebox{270}{\includegraphics[width=4cm,height=10cm]{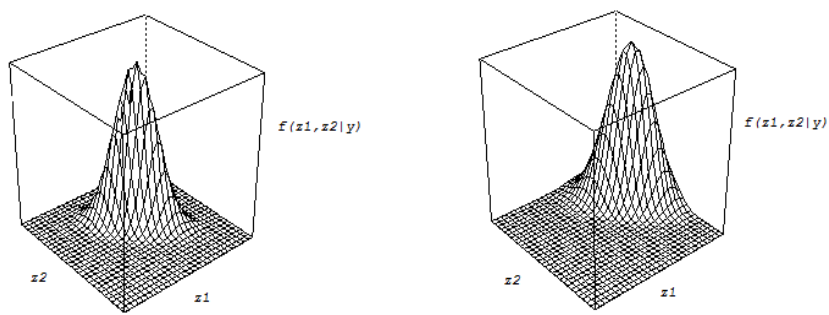}}\vspace{1cm}
\end{figure}
\end{center}

To investigate empirically the finite sample performance of the FLA
and AGH, based on both the posterior mean (AGH$_{me}$) and mode
(AGH$_{mo}$), we generated data from a population that consists of
five variables and satisfies a two factor model. The number of
categories is the same for each observed variable, and equal to 4.
100 random samples were considered with $n=200$ subjects. We chose 5
quadrature points per each latent variable for both the adaptive
approximations. We also considered 7 quadrature points, but there
was a little difference with 5 nodes, suggesting that the latter
provides sufficient accuracy for this example.\\ The population
parameters were chosen in such a way that the thresholds range from
-3 to 3. The factor loadings are the following: $\boldsymbol
\alpha_{1}=(1.03,1.44,2.11,1.8,1.53)$ and $\boldsymbol
\alpha_{2}=(0,2.42,1.52,0.75,1.34)$ with not null values generated
from a log-normal distribution, and one loading fixed to 0 to get a
unique solution. \\Table \ref{Tab1} reports the mean, bias, and Mean
Square Error (MSE) of the parameter estimates obtained by applying
all the techniques.
\begin{table}\begin{tiny}
\caption{Mean, bias and MSE of the parameter estimates for FLA,
AGH$_{me}$, and AGH$_{mo}$ in the generated data. \label{Tab1}}
\begin{tabular}{lccccccccccc} \hline
 & \multicolumn{7}{c}{\emph{AGH}} &&      \multicolumn{3}{c}{\emph{FLA}}  \\
\cline{2-8} \cline{10-12}
 & \multicolumn{3}{c}{mean} &&     \multicolumn{3}{c}{mode}    \\
 \cline{2-4} \cline{6-8} \\
\% valid samples   &\multicolumn{3}{c}{\emph{87}}&&\multicolumn{3}{c}{\emph{84}}&&\multicolumn{3}{c}{\emph{76}}\\\\
 True                 &  Mean  & Bias   & MSE  && Mean  & Bias  & MSE   && Mean &  Bias & MSE \\\\
$\alpha_{11}= 1.03$   &  1.52  & 0.49  & 0.96  &&1.52  & 0.49   & 0.68  && 1.05 &  0.02& 0.03 \\
$\alpha_{21}= 1.44$   &  1.48  &  0.04 & 0.53  &&1.20  &-0.24   & 0.38    && 1.37&-0.07 & 0.14\\
$\alpha_{31}= 2.11$   &  2.19  &  0.08 & 0.82  &&1.88  &-0.23   & 0.44  &&  2.08&-0.03 &  0.31\\
$\alpha_{41}= 1.80$   &  1.81  &  0.01 & 0.63  &&1.65  &-0.15   & 0.44  &&  1.44&-0.36& 0.21\\
$\alpha_{51}= 1.53$   &  1.53  &  0.00 & 0.42  &&1.36  &-0.17   & 0.27  &&  1.54& 0.01& 0.14\\
$\alpha_{12}= 0.00$   &   -    & -     & -     &&-     & -      & -       &&- & - & -  \\
$\alpha_{22}= 2.42$   &  2.04  & -0.38  & 0.60 &&2.10  & -0.33  & 0.41    && 2.04& -0.38& 0.35\\
$\alpha_{32}= 1.52$   &  1.62  & 0.10& 0.56   && 1.82  & 0.30   & 0.45  && 2.02& 0.50&0.51  \\
$\alpha_{42}= 0.75$   &  0.78  & 0.03 & 0.44  && 0.96  & 0.21  & 0.36   && 1.11& 0.36&  0.18\\
$\alpha_{52}= 1.34$   &  1.50  &0.16  & 0.38   &&1.63  & 0.29   & 0.41  && 1.76 & 0.42& 0.35\\
\hline
\end{tabular} \end{tiny}
\end{table}
The results show that the percentage of valid samples is quite high
for all the procedures, ranging from 76\% to 87\%. The FLA presents
much better MSE values than those achieved by AGH$_{me}$ and
AGH$_{mo}$, mainly due to a smaller variability of the estimates.
Comparing the adaptive techniques, AGH$_{me}$ estimates are less
biased than those determined by AGH$_{mo}$, and present an opposite
sign of the bias for $\boldsymbol \alpha_{1}$. On the other hand,
the latter behaves better in terms of MSE values.\\ The different
performance of the two adaptive techniques can be due to the fact
that the individual posterior densities to be approximated are not
always symmetric. In latent variable models for ordinal data,
\cite{Ch96} proved that the posterior densities asymptotically
follow a multivariate normal distribution. However, for a small
number of observed variables, the integrand could be skewed, and the
numerical procedures could provide quite different results.\\ To
analyze the shape of the individual posterior densities in the
generated population, we computed measures of multivariate skewness
$\beta_{1,q}$ and kurtosis $\beta_{2,q}$ proposed by Mardia (1970).
In the case of two latent variables, they are given by
$$\beta_{1,2}(l)=\mu_{30}^2+\mu_{03}^2+3\mu_{12}^2+3\mu_{21}^2 \quad
l=1,...,n$$ and $$\beta_{2,2}(l)=\mu_{04}+\mu_{40}+2\mu_{22} \qquad
l=1,...,n$$  with $\mu_{ij}=E(z_{1l}^{i}z_{2l}^j)$, whereas $z_{1l}$
and $z_{2l}$ are the latent factors standardized with respect to the
posterior densities. Mardia (1970) also derived the asymptotic
distributions of both $\beta_{1,q}$ and $\beta_{2,q}$, and the
corresponding statistical tests to evaluate the null hypotheses
$H_{0}:\beta_{1,q}=0$ and $H_{0}:\beta_{2,q}=q(q+2)$, being $q(q+2)$
the kurtosis in $q$-variate normal densities.\\ By computing these
measures for the individual posterior densities generated in this
simulation study, we observed that about 35\% of these  functions
have a significant skewness, and a kurtosis always not significantly
different from 8. In particular, $\beta_{1,2}$ is on average equal
to 0.044 and it ranges from 0.000 to the significant value 0.168.
The presence of individual posterior densities having different
shapes could justify the different behavior of AGHs and FLA. In
Figure \ref{Fig1}, we show two different functions obtained from our
generated data.
In order to better analyze the finite sample properties of FLA and
AGHs, we also generated data from  two hypothetical extreme
scenarios: one in which all the posterior densities are symmetric,
and another one in which a high percentage (more than 60\%) of the
densities are skewed.  As before, we consider five observed
variables, each with 4 categories, satisfying a two factor model.
The results for both the populations are shown in Table \ref{Tab2}.
\\In the first scenario, the thresholds for each item are  equal
to -2 for the first category, 0 for the second, and 2 for the third
one, whereas the loadings are all fixed to 0.5 except one set equal
to zero. In this population, all the individual posterior densities
are symmetric, with $\beta_{1,2}$ on average equal to 0.005, and
$\beta_{2,2}$ always not significantly different from 8. As in the
previous simulation study, we generated 100 random samples with 200
subjects.\\For all the samples the algorithm achieves the
convergence for FLA and AGH$_{me}$, and in the 96\% of the cases for
AGH$_{mo}$. The FLA  improves a lot with respect to the previous
case, with a reduction of almost one digit in the MSE values, mainly
due to smaller bias values for $\boldsymbol \alpha_{2}$. On the
other hand, both the AGH techniques provide better results in terms
of bias and MSE, even if they still perform worse than FLA. We can
also notice that the results provided by the two adaptive procedures
are almost the same,
with an equal sign of the bias for all the estimates,
and  slight discrepancies due to the different computational techniques involved.
 Indeed, as discussed by \cite{HesSkPi05}, the two procedures should provide similar results when the posterior densities are symmetric.\\
In the second scenario, the thresholds for each item are equal to -1 for the first category, 0 for the second, and 1 for the third one, whereas the loadings are fixed equal to $\boldsymbol \alpha_{1}=(2.5,2.5,2.5,2.5,2.5)$ and $\boldsymbol \alpha_{2}=(0,1,1,1,1)$. In this case, the 65\% of the  posterior densities are skewed.
$\beta_{1,2}$ ranges from 0.000 to 0.239, being the latter significantly different from zero, and it is on average equal to 0.127. On the other hand, there is not significant kurtosis for all the subjects.
The main consequence of this high percentage of skew densities is that, for both FLA and AGH$_{mo}$,
a very small number of samples (27\% for the former, 35\% for the latter) converge properly. Hence, even if the results are similar to the ones obtained in the first simulation, they are not reliable.
On the other hand, AGH$_{me}$ seems to be not affected by the different shapes of the posterior densities. It results more stable in terms of mean, bias, and MSE of the estimates as well as in terms of percentage of valid samples, that also in this case is 83\%.\\
\begin{table}
\begin{tiny}
\caption{Mean, bias and MSE for FLA, AGH$_{me}$, and AGH$_{mo}$ for
different scenarios in finite samples. \label{Tab2}}
\begin{tabular}{lccccccccc} \hline\\
& \multicolumn{6}{c}{\emph{AGH}} &      \multicolumn{3}{c}{\emph{FLA}}  \\
&\cline{1-6}
 & \multicolumn{3}{c}{mean} &     \multicolumn{3}{c}{mode}  &&& \\
&\cline{1-6}\\

\% valid samples   &\multicolumn{3}{c}{\emph{100}}&\multicolumn{3}{c}{\emph{96}}&\multicolumn{3}{c}{\emph{100}}\\\\

True                &  Mean  & Bias   & MSE  & Mean  & Bias  & MSE   & Mean &  Bias & MSE \\
$\alpha_{11}=0.5$   &  0.78  & 0.28  & 0.59  & 0.79  & 0.29   & 0.62   & 0.59 &  0.09& 0.02    \\
$\alpha_{21}=0.5$   &  0.46  & -0.04 & 0.39  & 0.50  &-0.00   & 0.46       & 0.61& 0.11 & 0.02    \\
$\alpha_{31}=0.5$   &  0.43  &  -0.07& 0.50  & 0.41  &-0.09   & 0.48   &  0.60& 0.10 &  0.02   \\
$\alpha_{41}=0.5$   &  0.52  &  0.02 & 0.78  & 0.54  & 0.04   & 0.77   &  0.60& 0.10& 0.02     \\
$\alpha_{51}=0.5$   &  0.61  &  0.11 & 0.67  & 0.60  & 0.10   & 0.64   &  0.61& 0.11& 0.02     \\
$\alpha_{12}=0.0$   &   -    & -     & -     & -       & -      & -        &- & - & -             \\
$\alpha_{22}=0.5$   &  0.81  &  0.31  & 0.80 & 0.83  &  0.33  & 0.82       & 0.60&  0.10& 0.01     \\
$\alpha_{32}=0.5$   &  0.64  & 0.14& 0.79   &  0.58  & 0.08   & 0.66   & 0.59& 0.09&0.01        \\
$\alpha_{42}=0.5$   &  0.80  & 0.30 & 0.77  &  0.77  & 0.27  & 0.85    & 0.59& 0.09&  0.01      \\
$\alpha_{52}=0.5$   &  0.72  &0.22  & 0.61    &0.74  & 0.24   & 0.65   & 0.59 & 0.09& 0.01      \\
\hline
\hline\\
\hline\\
& \multicolumn{6}{c}{\emph{AGH}} &      \multicolumn{3}{c}{\emph{FLA}}  \\
&\cline{1-6}
 & \multicolumn{3}{c}{mean} &     \multicolumn{3}{c}{mode}  &&& \\
&\cline{1-6}\\

\% valid samples   &\multicolumn{3}{c}{\emph{83}}&\multicolumn{3}{c}{\emph{27}}&\multicolumn{3}{c}{\emph{35}}\\\\

True                &  Mean  & Bias   & MSE  & Mean  & Bias  & MSE   & Mean &  Bias & MSE \\
$\alpha_{11}=2.5$   &  2.49  &-0.01  & 0.31  & 2.35  &-0.16   & 0.18   & 1.82 & -0.68& 0.60    \\
$\alpha_{21}=2.5$   &  2.74  & 0.24 & 0.37  & 2.56  &0.06   & 0.15     & 2.42&-0.08 & 0.19    \\
$\alpha_{31}=2.5$   &  2.79  & 0.29& 0.58  & 2.47  &-0.03   & 0.16   &  2.36&-0.14 &  0.16   \\
$\alpha_{41}=2.5$   &  2.63  &  0.13 & 0.24  & 2.57  & 0.07   & 0.12   &  2.41&-0.09& 0.08     \\
$\alpha_{51}=2.5$   &  2.75  &  0.25 & 0.41  & 2.61  & 0.11   & 0.14   &  2.37&-0.13& 0.15     \\
$\alpha_{12}=0.0$   &   -    & -     & -     & -       & -      & -        &- & - & -             \\
$\alpha_{22}=1.0$   &  1.04  &  0.04  & 0.44 & 0.90  & -0.10  & 0.35       & 1.86&  0.86& 0.81     \\
$\alpha_{32}=1.0$   &  1.26  & 0.26& 0.52   &  0.94  &-0.07   & 0.34   & 1.82& 0.82&0.76        \\
$\alpha_{42}=1.0$   &  1.07  & 0.07 & 0.50  &  1.09  & 0.09  & 0.65    & 1.90& 0.90&  0.87      \\
$\alpha_{52}=1.0$   &  1.23  &0.23  & 0.65    &0.94  &-0.06   & 0.61   & 1.87 & 0.87& 0.80      \\
\hline
\hline\\
\end{tabular} \end{tiny}
\end{table}From these results, we can argue that FLA  will be superior than AGH
when the majority of the posterior densities is  symmetric. In these
cases the former provides better MSE values for the estimates than
the latter, mainly due to a reduced variability in the estimates.
Moreover, the bias introduced in the estimates using FLA is quite
comparable with the one in the AGH estimates. On the other hand, we have also shown that
the AGH$_{me}$ provides more stable results, that are not affected by the
shape of the integrand.  Its use is then suggested in populations
characterized by a high percentage of skew distributions.

\subsection{Asymptotic properties of estimators}

The asymptotic properties of the Laplace maximum likelihood estimators $\hat{\theta}$ have been derived and discussed by \cite{RizVer09}. Under suitable regularity conditions, these authors showed that $$\hat{\theta}-\theta_{0}=O_{p}\left[\max\left\{n^{-1/2}, p^{-2}\right\}\right],$$ where $\theta_{0}$ denotes the true parameter value.  $\hat{\theta}$ will be consistent as long as both $n$ and $p$ grow to $\infty$. 
FLA is superior than standard Laplace method, the latter producing
estimators with an approximation error of order
$O_{p}\left[\max\left\{n^{-1/2}, p^{-1}\right\}\right]$. On the
other hand, following \cite{LiuPie94}
and \cite{TieKa89}, it can be shown that FLA shares the same approximation error of the AGH with 5 quadrature points.\\To assess
 the asymptotic accuracy of the FLA estimators, we generated 100 random samples with 1000 subjects from the population described in the previous section. We also applied both the adaptive techniques, and the results are shown in Table \ref{Tab3}.\\
\begin{table}\begin{tiny}
\caption{Mean, bias and MSE for FLA, AGH$_{me}$, and AGH$_{mo}$ for
generated data with $n=1000$.\label{Tab3}}
\begin{tabular}{lccccccccccc} \hline
 & \multicolumn{7}{c}{\emph{AGH}} &&      \multicolumn{3}{c}{\emph{FLA}}  \\
\cline{2-8} \cline{10-12}
 & \multicolumn{3}{c}{mean      } &&     \multicolumn{3}{c}{mode}    \\
 \cline{2-4} \cline{6-8}\\
\% valid samples   &\multicolumn{3}{c}{\emph{97}}&&\multicolumn{3}{c}{\emph{92}}&&\multicolumn{3}{c}{\emph{99}}\\\\
 True                 &  Mean  & Bias   & MSE  && Mean  & Bias  & MSE   && Mean &  Bias & MSE \\\\
$\alpha_{11}= 1.03$   &  1.11  & 0.08  & 0.11  && 1.13 &  0.10  & 0.09  && 1.04& 0.01& 0.01     \\
$\alpha_{21}= 1.44$   &  1.43  & -0.01& 0.10  && 1.39  & -0.05 & 0.10  && 1.39 &-0.05& 0.04\\
$\alpha_{31}= 2.11$   &  2.12  &  0.01 & 0.09  && 2.09  & -0.02 & 0.09  && 2.15& 0.04& 0.07\\
$\alpha_{41}= 1.80$   &  1.85  & 0.05 & 0.12  && 1.83  & 0.03 & 0.13  && 1.50&  0.05& 0.04   \\
$\alpha_{51}= 1.53$   &  1.55  &  0.02 & 0.06  && 1.52  & -0.01 & 0.06  && 1.59&  0.05& 0.04\\
$\alpha_{12}= 0.00$   &   -    & -     & -     && -     &  -    & -     &&- & - & -  \\
$\alpha_{22}= 2.42$   &  2.17  & -0.25  & 0.20 && 2.23  & -0.19 & 0.19  && 2.16& -0.26& 0.15\\
$\alpha_{32}= 1.52$   &  1.56  & 0.04& 0.11   && 1.61  &  0.09 & 0.12  && 2.04& 0.52&  0.32  \\
$\alpha_{42}= 0.75$   &  0.74  &-0.01 & 0.09  && 0.78  &   0.03 & 0.09   && 1.05 &  0.30 &  0.10 \\
$\alpha_{52}= 1.34$   &  1.39  &0.05  & 0.07   && 1.42  &  0.08 & 0.07  && 1.73& 0.39&  0.20 \\
\hline
\end{tabular}
\end{tiny}
\end{table}
The percentage of valid samples is high for all the techniques, ranging from 92\% to 99\%.
FLA has a good performance as before with small MSE and bias values.
On the other hand, both AGHs have an analogous behavior: the MSE
values are drastically reduced with respect to the finite sample
situation, and the bias is small for all the parameters.\\ The three
techniques present a very similar asymptotic behavior. Moreover, it is worth
noting that FLA performs better than the classical Laplace
approximation. Indeed, the asymptotic bias
of the latter is higher than the one corresponding to AGH \citep{Joe08},
whereas the bias in the AGH and FLA estimates is quite
comparable, with a slight better performance of the former for the
second factor loadings (Table \ref{Tab3}).

\section{Discussion}

This paper is concerned with the adequacy of several approximations
of the likelihood function in latent variable models for ordinal
data. In particular, we proposed an extended version of the Laplace
method for approximating integrals, known as fully exponential
Laplace approximation. Classical Laplace methods are known to work
poorly in presence of discrete response variables \citep{Joe08}, but we have shown how the FLA is generally
appropriate in models for ordinal data in  both finite and large
samples. The comparison with the adaptive Gauss Hermite quadrature
techniques has highlighted that in finite samples the FLA provides
better results in terms of MSE values when the majority of the
posterior densities is symmetric. Indeed, for a small number of
observed variables, the symmetry  of the individual posterior
densities is not always guaranteed, and the percentage of skew
distributions tends to vary according to the parameter values. When
the majority of the densities are skewed, FLA and AGH$_{mo}$ do not
achieve converge in most cases. On the other hand, AGH$_{me}$ is
more stable, and it is not affected by the shape of the functions to
be approximated.  \\
The main strength of the FLA approach is that it effectively copes
with high dimensional latent structures without increasing
substantially the computational burden.  This is one of the main
drawbacks in the application of AGH techniques in latent variable
models. Five quadrature points can provide accurate estimates, but
the computational effort increases exponentially as the number of
latent factors increases. Furthermore, in large samples, the FLA
achieves the same approximation of the AGH with
five quadrature points, and all the techniques behave similarly.\\
The main limitation of the FLA approach is that it is not possible
to control the magnitude of the approximation error of the integral,
as done in AGH by modifying the number of quadrature points.
However, as discussed by \cite{RizVer09}, a virtue of the fully
exponential Laplace approximation is that it is very general, and it
can be used in almost all the general linear latent variable
models. Overall, for latent variable models with ordinal data, the
FLA is very adequate to approximate the likelihood function, and it
should be considered as a valid alternative to adaptive Gaussian
quadrature
techniques. \\
Further lines of research will be oriented to compare the
performance of  FLA with the multidimesional splines. The latter
represents a useful alternative to approximate the posterior
densities \citep{WoTie06} and to investigate the main assumptions on
the prior distribution of the latent variables that is still an open
issue in the GLLVM framework \citep{KnoTza07}.

\bibliographystyle{plainnat}
\bibliography{BibLap}

\section*{Appendix}
\noindent In order to apply the fully exponential Laplace approximation, the first derivative of $\boldsymbol \Sigma_{l}$ with respect to $t$ has to be computed. At this regard, we make use of the following result
$$\frac{\partial\boldsymbol\Sigma_{l}^{(t)}}{\partial t}\Big|_{(\mathbf{z}_{l}=\mathbf{\hat{z}}_{l}, t=0)}=\frac{\partial\boldsymbol\Sigma_{l}^{(t)}}{\partial \mathbf{z}}\frac{\partial \mathbf{\hat{z}}_{l}}{\partial t}\Big|_{(t=0)}$$
according to which
\begin{eqnarray*}
\frac{\partial\boldsymbol\Sigma_{l}^{(t)}}{\partial t}\Big|_{(\mathbf{z}_{l}=\mathbf{\hat{z}}_{l}, t=0)} &=& \sum_{i=1}^{p}\sum_{s=1}^{c_{i}-1}\boldsymbol \alpha_{i}\boldsymbol \alpha'_{i}[y_{i,s,l}\gamma_{i,s+1,l}(1-3\gamma_{i,s+1,l}+2\gamma_{i,s+1,l}^{2})+\\
&-&y_{i,s+1,l}\gamma_{i,s,l}(1-3\gamma_{i,s,l}+2\gamma_{i,s,l}^{2})] \times \boldsymbol\alpha_{i}\Sigma_{l}^{-1}A'(\mathbf{\hat{z}}_{l})-A''(\mathbf{\hat{z}}_{l}),
\end{eqnarray*}
where $A'(\mathbf{\hat{z}}_{l})=\frac{\partial A(\mathbf{z}_{l})}{\partial \mathbf{z}_{l}}\Big|_{\mathbf{{z}}_{l}=\mathbf{\hat{z}}_{l}}$ and $A''(\mathbf{\hat{z}}_{l})=\frac{\partial^{2} A(\mathbf{z}_{l})}{\partial \mathbf{z}'_{l}\partial \mathbf{z}_{l}}\Big|_{\mathbf{{z}}_{l}=\mathbf{\hat{z}}_{l}}$.\\ For the thresholds, the first-order partial derivatives result
$$A'_{1,i,s}(\mathbf{z}_{l})=\boldsymbol \alpha_{i}\gamma_{i,s,l}(1-\gamma_{i,s,l}), \quad s=1,\ldots,c_{i}-1$$
and
$$A'_{2,i,s}(\mathbf{z}_{l})=-\boldsymbol \alpha_{i}\gamma_{i,s,l}(1-\gamma_{i,s,l}) \quad s=1,\ldots,c_{i}-1.$$
for $A_{1,i,s}(\boldsymbol z_{l})$ and $A_{2,i,s}(\boldsymbol z_{l})$, respectively. Furthermore, the corresponding second-order partial derivatives are given by
$$A''_{1,i,s}(\mathbf{z}_{l})=-\boldsymbol \alpha_{i}'\boldsymbol \alpha_{i}\gamma_{i,s,l}(1-3\gamma_{i,s,l}+2\gamma_{i,s,l}^{2}), \quad s=1,\ldots,c_{i}-1$$
and
$$A''_{2,i,s}(\mathbf{z}_{l})=\boldsymbol \alpha_{i}'\boldsymbol \alpha_{i}\gamma_{i,s,l}(1-3\gamma_{i,s,l}+2\gamma_{i,s,l}^{2}) \quad s=1,\ldots,c_{i}-1.$$
As for the loadings, the elements of the gradient $A_{3,i,s}(\boldsymbol z_{l})$ with respect to the latent variables result\\

\noindent $\frac{\partial A_{3,i,s}(z_{jl})}{\partial z_{jl}}=\left\{\begin{array}{ll}
                                    (1-\gamma_{i,1,l})(1+\gamma_{i,1,l}\alpha_{ij}z_{jl}), & \hbox{if $s=1$;} \\
                                    (1-\gamma_{i,s,l}-\gamma_{i,s-1,l})+\alpha_{ij}z_{jl}(\gamma_{i,s,l}(1-\gamma_{i,s,l})+\gamma_{i,s-1,l}(1-\gamma_{i,s-,l})), & \hbox{if $s=2,\ldots,ci$.} \\
                                    \end{array}
                                   \right.
$\\

\noindent $\frac{\partial A_{3,i,s}(z_{jl})}{\partial z_{kl}}=\left\{\begin{array}{ll}
                                    \alpha_{ik}z_{jl}\gamma_{i,1,l}(1-\gamma_{i,1,l}), & \hbox{if $s=1$;} \\
                                     \alpha_{ik}z_{jl}(\gamma_{i,s,l}(1-\gamma_{i,s,l})+\gamma_{i,s-1,l}(1-\gamma_{i,s-,l})), & \hbox{if $s=2,\ldots,ci$} \\
                                    \end{array}
                                   \right.
$\\

\noindent On the other hand, the elements of the corresponding Hessian matrix are given by\\

\noindent $\frac{\partial^{2} A_{3,i,s}(z_{jl})}{\partial z_{jl}^{2}}=\left\{\begin{array}{ll}
                                    \alpha_{ij}\gamma_{i,1,l}(1-\gamma_{i,1,l})(2-\alpha_{ij}z_{jl}(1-2\gamma_{i,1,l})), & \hbox{if $s=1$;} \\
                                    \alpha_{ij}[\gamma_{i,s,l}(1-\gamma_{i,s,l})(2-\alpha_{ij}z_{jl}(1-2\gamma_{i,s,l}))+ & \hbox{if $s=2,\ldots,ci$.} \\
                                    +\gamma_{i,s-1,l}(1-\gamma_{i,s-1,l})(2-\alpha_{ij}z_{jl}(1-2\gamma_{i,s-1,l}))],&\\
                                    \end{array}
                                   \right.
$\\

\noindent $\frac{\partial^{2} A_{3,i,s}(z_{jl})}{\partial z_{jl} \partial z_{kl}}=\left\{\begin{array}{ll}
                                    \alpha_{ik}\gamma_{i,1,l}(1-\gamma_{i,1,l})(1-\alpha_{ik}z_{jl}(1-2\gamma_{i,1,l})), & \hbox{if $s=1$;} \\
                                    \alpha_{ik}[\gamma_{i,s,l}(1-\gamma_{i,s,l})(1-\alpha_{ij}z_{jl}(1-2\gamma_{i,s,l}))+ & \hbox{if $s=2,\ldots,ci$.} \\
                                    +\gamma_{i,s-1,l}(1-\gamma_{i,s-1,l})(1-\alpha_{ij}z_{jl}(1-2\gamma_{i,s-1,l}))],&\\
                                    \end{array}
                                   \right.
$\\

\noindent $\frac{\partial^{2} A_{3,i,s}(z_{jl})}{\partial z_{kl}^{2}}=\left\{\begin{array}{ll}
                                    -\alpha_{ik}^2z_{jl}\gamma_{i,1,l}(1-3\gamma_{i,1,l}+2\gamma_{i,1,l}^{2}), & \hbox{if $s=1$;} \\
                                    -\alpha_{ik}^{2}z_{j}(\gamma_{i,s-1,l}-3\gamma_{i,s-1,l}^{2}+2\gamma_{i,s-1,l}^{3}+\gamma_{i,s,l}-3\gamma_{i,s,l}^{2}+2\gamma_{i,s,l}^{3}), & \hbox{if $s=2,\ldots,ci$.} \\
                                    \end{array}
                                   \right.
$\\

\end{document}